\documentclass[12pt]{article}
\usepackage[dvips]{graphicx}
\def\S{\Sigma}
\def\R{R}
\def\D{{\cal D}}
\def\<{\langle}
\def\>{\rangle}
\def\tr{{\rm trace}}
\def\det{{\rm det}}
\def\moins{-{1\over2}}
\def\plus{+{1\over2}}

\def\Pv{{\bf P}}
\def\Sv{{\bf S}}
\def\pv{{\bf p}}
\def\kv{{\bf k}}
\def\nv{\hat{\bf n}}
\def\xu{\hat{\bf x}}
\def\yu{\hat{\bf y}}
\def\zu{\hat{\bf z}}

\newcommand{\ppLL}{\ensuremath{\bar{\rm p}+{\rm
p}\to\overline{\Lambda}+\Lambda}}
\catcode`\@=11
\renewcommand\section{\@startsection {section}{1}{\z@}%
                                  {-3.5ex \@plus -1ex \@minus -.2ex}%
                                  {2.3ex \@plus.2ex}
{\noindent \bf    }}
\let\theeq=\theequation
\renewcommand{\theequation}{\thesection.\theeq}
\catcode`\@=12
\def\be{\begin{equation}}
\def\ee{\end{equation}}
\catcode`\@=11
\def\eqalign#1{\null\,\vcenter{\openup\jot\m@th
\ialign{\strut\hfil$\displaystyle{##}$&$\displaystyle{{}##}$\hfil
    \crcr#1\crcr}}\,}
\catcode`\@=12
\begin{document}
\begin{flushright}\begin{small}\begin{obeylines}
Preprint LPSC 03-67 
 LYCEN 2004-04
 hep-ph/0401234
\end{obeylines}\end{small}
\end{flushright}

\vskip .4cm
\begin{center}
{\bfseries GENERAL CONSTRAINTS ON SPIN OBSERVABLES;\\
APPLICATIONS TO  \boldmath\ppLL\unboldmath\\
AND TO POLARIZED QUARK DISTRIBUTIONS}%
\footnote{Invited talk by X.~Artru at the Xth Advanced Research  Workshop on High Energy Spin Physics (DUBNA-SPIN-03),
Sept. 16-20, Dubna}
\vskip 5mm
{X. Artru}$^{\dag}$ and J.-M. Richard$^{ \ddag }$
\vskip 5mm
\begin{small}
${}^{\dag }\,${\it Institut de Physique Nucl\'eaire de Lyon\\
IN2P3-CNRS and Universit\'e Claude Bernard,
F-69622 Villeurbanne, France} \\
{\it E-mail: x.artru@ipnl.in2p3.fr} \\
${}^{\ddag}\,${\it Laboratoire de Physique Subatomique et Cosmologie\\
Universit\'e Joseph Fourier and IN2P3-CNRS,
F-38026 Grenoble, France}\\
{\it E-mail: jean-marc.richard@lpsc.in2p3.fr}
\end{small}
\end{center}
\vskip 5mm

\begin{abstract}
We review the density matrix formalism and the positivity conditions for general
multiple spin asymmetries, taking as an example the case \ppLL\
in which one, two or three spins are analyzed.
Some aspects related to quantum information and entangled states are discussed.
Some positivity domains for pairs and triplets of spin parameters are displayed,
together with the experimental points. The case of inclusive reaction is also
treated, taking as an example the spin- and transverse momentum- dependent
quark distributions.
\end{abstract}

\vskip 8mm
\setcounter{equation}{0}
%
\noindent
{\bf 1. Introduction}
\medskip

The single- or multiple-spin asymmetries which can be measured using polarized beams,
polarized targets or analyzing the final-particle polarizations
provide important information about the elementary processes in particle
physics.
These spin observables may be related by {\it equalities} coming from the symmetries
of the processes.
Besides, they satisfy {\it inequalities}
expressing the positivity of a {\it Grand Density Matrix}, $\R$,
which describes all possible polarized cross sections.
The positivity of $\R$ insures that the cross section is positive
for any initial and final spin states, including entangled ones.
The resulting inequalities provide consistency checks of experimental data,
constrain any parametrization of polarized structure functions
and, in the future, may be applied to Monte-Carlo event generators with spins.

The positivity conditions are also interesting
from the point of view of  the {\it quantum information} carried by spins.
The information about the scattering amplitudes is maximal
when all the independent spin observables are measured.
There are inequalities that define the allowed domain and are saturated in
this case.
Conversely, there is a loss of information, in other words an increase of entropy,
when some particles are not analyzed or in the case of inclusive reactions.
In this case, most of the inequalities may be non-saturated.

There is an abundant literature on positivity conditions. Some key papers of the
60's are still relevant. See, for instance, Ref.~\cite{Leader} for a survey and
some references.  The subject has been revisited in recent months due to
the results on the reaction
$$
\ppLL~, \eqno(1.1)
$$
measured by the PS185 collaboration at CERN
\cite{Klempt}. The polarization of the outgoing hyperon or antihyperon is
revealed by its weak decay. Since a polarized target was used in the last
runs, observables up to rank 3 can be accessed.

This contribution is
a short introduction to the density matrix formalism
and derivation of the positivity conditions.
In Sections 2-5 we will consider the case of {\it exclusive} reactions,
illutrating it by reaction (1.1).
We will also present in Section 6 some results
of an alternative ``empirical" method
by which the positivity domains of a subset of observables
can be very easily discovered.

Finally, we will consider the case of {\it inclusive} reactions
and obtain the inequalities which must be satisfied by the
{\it spin-dependent quark distribution},
considered as the probability of the elementary splitting process
$$
{\rm nucleon} \to {\rm quark}  + X \,.
\eqno(1.2)
$$
for a given longitudinal momentum ratio
$ x = p_L ({\rm quark}) / p_L ({\rm nucleon}) $.

\bigskip
\noindent
{\bf 2. The spin observables}
\medskip

The fully polarized differential cross section of (1.1),
more generally $A+B \to C+D$, where $A$, $B$, $C$ and $D$ have spin ${1\over2}$,
can be expressed as
$$
{d\sigma \over d\Omega} \left( \Sv_A,\Sv_B,\Sv_C,\Sv_D \right)
= {1\over4} \, \left( {d\sigma\over d\Omega} \right)_{\rm unpol}  \
C_{\lambda\mu\nu\tau} \ S^\lambda_A \, S^\mu_B \, S^\nu_C \, S^\tau_D
\ .
\eqno(2.1)
$$
The $\Sv\,$'s are the polarization vectors of pure spin states ($|\Sv| = 1$).
In the right-hand side they are promoted to four-vectors with $S^{0} = 1$.
The indices $\lambda, \mu, \nu, \tau$, run from 0 to 3,
whereas latin indices $i$, $j$, $k$, $l$, take the values 1, 2, 3, or $x$, $y$, $z$.
A summation is understood over each repeated index.
$S^x, S^y, S^z$ are measured in a triad of unit vectors
$ \{ \xu, \yu, \zu \} $ which may differ from one particle to the other.
$C_{\lambda\mu\nu\tau}$ are the {\it correlation parameters}.
For example,
$C_{0000} \equiv 1$,
$C_{xy00} \equiv A_{xy}$ is an initial double-spin asymmetry,
$C_{000y}$ is the spontaneous polarization of particle $D$ along $\yu$,
$C_{0y0y} \equiv D_{yy}$ is a spin transmission coefficient from $B$ to $D$ and
$C_{00xy} \equiv C_{xy}$ is a final spin correlation.

Equation (2.1) also applies to the case of incomplete initial polarizations,
replacing the unit vector $\Sv_A$ by $\Pv_A$ with $|\Pv_A| \le 1$
and the same for $B$.
The final polarizations generally depend on the initial ones, e.g.,
$$
\Pv_C \equiv \< \Sv_C \>  =
\left( {1\over4} \, {d\sigma\over d\Omega} \right)_{\rm unpol}^{-1}  \
\nabla_{ \Sv_C } \ {d\sigma \over d\Omega}
\left( \Pv_A,\Pv_B,\Sv_C \, ; \, \Sv_D = 0 \right) ~.
\eqno(2.2)
$$

\goodbreak\goodbreak\bigskip
\noindent
{\bf 3. The density matrix formalism}
\medskip

Let $ \{ | \alpha \> \} $, with $ \alpha = \pm {1\over2}$,
be the basic spin states of particle $A$,
$ \{ | \beta \> \} $ those of $B$, etc.
The quantification axis $\zu$ may differ from one particle to the other.
It can be the {\it helicity} axis $\pv / |\pv|$
or the {\it transversity} axis,
$ \nv = \pv_A \times \pv_C \, / \, | \pv_A \times \pv_C | $.
We write the spin-dependent amplitude of (1.1) as
$$
( \, \< \bar\Lambda, \gamma | \otimes \< \Lambda, \delta| \, )
\ M \
( \, |\bar p, \alpha \> \otimes | p,  \beta \> \, )
\equiv \< \gamma  \delta |  M | \alpha  \beta \>~.
\eqno(3.1)
$$
For each spin ${1\over2}$ particle we have the single-spin density matrix
$$
\rho(\Pv) \equiv {1\over2} (1 + \Pv\cdot \sigma) \,,
\eqno(3.2)
$$
For {\it partial} polarizations of $A$ and $B$, but definite spins of $C$ and $D$,
the cross-section (2.1) becomes
$$
{d\sigma \over d\Omega} \left( \Pv_A,\Pv_B,\Sv_C,\Sv_D \right) =
\tr \{ \, M  \ \rho(\Pv_A) \otimes  \rho(\Pv_B)  \, M^\dagger
\rho(\Sv_C) \otimes \rho(\Sv_D)  \}~.
\eqno(3.3)
$$
The final two-spin density matrix is
$$
\rho_{C,D} = { M \ \rho(\Pv_A) \otimes  \rho(\Pv_B)  \ M^\dagger
\over \tr \{ M \, M^\dagger \} }~.
\eqno(3.4)
$$
It describes the individual polarizations of $C$ and $D$ and their spin correlations.
The polarization of the $C$ is obtained
by taking the partial trace over the $D$ spin variable~:
$$
\rho_C  = \tr_{D} \ \rho_{C,D}\,,
\quad \hbox{i.e.,} \quad
\<\gamma | \rho_C |\gamma' \> = \sum_\delta
\<\gamma\delta | \rho_{C,D}  |\gamma'\delta \>~.
\eqno(3.5)
$$

The lack of information about a system can be measured
by various estimators, among which the entropy
$ S = - \tr \{\rho \log \rho\} $ and the rank of $\rho$.
Pure states (maximum  information) have zero entropy and unit rank.
The entropy (resp.\ rank) of the initial state is the sum (resp.\ product)
of the single-particle entropies (resp.\ ranks).
The rank of the final density matrix (3.4) is less than or equal to the initial one.
Therefore, complete initial polarizations
(~$| \Pv({\rm p})|= |\Pv(\bar{\rm p}) |=1)$ lead to a final pure state.
It does not imply that the individual polarization of the $\bar\Lambda$,
obtained from (3.5), is complete,
because the $ \bar\Lambda \Lambda $ state may be {\it entangled}
~\cite{Dolomieu}.

Let us now generalize the density matrix  in order to describe
in an unified way the spin {\it correlations} inside the final state
and the {\it transmission} of polarizations (i.e., of spin information)
between the initial and the final particles.
For this purpose we consider the fictitious {\it crossed} reaction of (1.1),
$$
|{\rm vacuum} \> \to  \rm{p} +\bar {\rm p} + \bar\Lambda+ \Lambda ~.
\eqno(3.6)
$$
We restrict this crossing to {\it spin} and {\it flavor} variables
(we do not consider the momenta). An initial particle {\it ket} becomes
a {\it final anti-particle bra} of {\it opposite spin}, for instance,
$$
| {\rm p}, \beta \>  \quad \to \quad \ \<  \bar{\rm p}, - \beta |
\equiv  \<  {\rm p}, \beta | \ CPT~, \eqno(3.7)
$$
where $C$, $P$ and $T$ are the charge conjugation,
parity and time-reversal operators
(it does not matter if reaction (3.6) does not conserve energy-momentum).
Accordingly, we can rewrite the spin-dependent amplitude (3.1) as
$$
\< \gamma,  \delta |  M | \alpha,  \beta \>
= \< -\alpha,  -\beta, \gamma,  \delta |
\, M^{\rm crossed} \, | {\rm vacuum} \>
\equiv \< -\alpha,  -\beta, \gamma,  \delta \, | \Psi \>~.
\eqno(3.8)
$$
Thus we produce a one-to-one correspondance
between the 2-particle transition {\it operator} $M$
and a 4-particle {\it state vector} $| \Psi  \>$,
which we will call the {\it Grand Wave Function}.
To shorten the equations, we will introduce the notation
$\bar\alpha \equiv -\alpha$, $\bar\beta \equiv  -\beta$, etc.
For explicit values of $\alpha$, we will use the notations $u$ and $d$
(for ``up" and ``down", like for quark isospin states)
instead of $\plus$ and $\moins$.
Therefore (3.7) will be written
$$
|u\> \to \< \bar{u}| , \quad |d\> \to \< \bar d | \,.
\eqno(3.9)
$$
The {\it Grand Density Matrix}, $\R$, which describes all possible spin correlations
in reaction (1.1), is defined by
$$
\< \bar\alpha \bar\beta \gamma \delta | \R
| \bar\alpha' \bar\beta' \gamma' \delta' \>
\ \equiv \
\< \gamma  \delta |  M | \alpha  \beta \>
\ \< \alpha' \beta' | M^\dagger | \gamma' \delta' \>
\ = \
\< \bar\alpha \bar\beta \gamma \delta | \Psi \>
\ \< \Psi | \bar\alpha' \bar\beta' \gamma' \delta' \> ~.
\eqno(3.10)
$$
Like ordinary density matrices, it is hermitian and semi-positive. Its trace is
given by
$$
\tr(\R) = \< \Psi | \Psi \> = \tr(M  M^\dagger)~,
\eqno(3.11)
$$
which is $2^{2}$ times the unpolarized cross section.
Dividing by (3.11), we can re-scale $\R$ to unit trace,
as a standard density matrix.
As can be seen from (3.10), $\R$ describes a pure state:
$\R  =  | \Psi \>    \< \Psi | $,
and is therefore of rank one.
This is a particular property of {\it exclusive} reactions.

The expression (3.4) for the final density matrix can be re-written as
$$
\rho(\bar\Lambda, \Lambda)
= { \tr_{\bar\alpha, \bar\beta}
\{ \  \R \ [ \rho^t(p) \otimes \rho^t(\bar p) ] \ \}
\over \tr (R) } \,,
\eqno(3.12)
$$
where $\rho^t(p)$ is the transpose of $\rho(p)$.
This transposition is explained in Appendix A.

The Grand Density Matrix can be expressed in terms of the correlation
parameters and vice-versa through
$$
\R = 2^{-4} \ C_{\lambda\mu\nu\tau} \ \sigma^t_\lambda(A) \otimes
\sigma^t_\mu(B) \otimes \sigma_\nu(C) \otimes\ \sigma_\tau(D)
\ ,
\eqno(3.13)
$$
$$
C_{\lambda\mu\nu\tau} = \tr \{ \, \R \ \left[
\sigma^t_\lambda(A) \otimes
\sigma^t_\mu(B) \otimes \sigma_\nu(C) \otimes\ \sigma_\tau(D)
\right]  \} \ ,
\eqno(3.14)
$$
where $\sigma_0$ is the unit $2\times2$ matrix.

\bigskip
\noindent
{\bf 4. Reduction of the density matrix}
\medskip

It is difficult to have polarized anti-protons.
Therefore the practical spin observables
in reaction (1.1) concern only $p$, $\Lambda$ and $\bar\Lambda$.
They are encoded in the sub-density matrix
$\R({\rm p}, \bar\Lambda, \Lambda) = \tr_{\bar\alpha}
\{\rho^t(\bar{\rm p}) \ \R(\bar{\rm p},{\rm p}, \bar\Lambda, \Lambda) \}$
with $\rho(\bar p)= {1\over2} I$, more explicitely
$$
\< \bar\beta \gamma \delta |
\ \R({\rm p}, \bar\Lambda, \Lambda) \
| \bar\beta' \gamma' \delta' \>
= \sum_{\bar\alpha} \< \bar\alpha \bar\beta \gamma \delta |
\ \R(\bar{\rm p}, {\rm p}, \bar\Lambda, \Lambda) \
| \bar\alpha \bar\beta' \gamma' \delta' \>
\ .
\eqno(4.1)
$$
This density matrix has dimension $8 \times 8$,
which is still rather large to write down the positivity conditions
(the original one was $16 \times 16$).
It has a non-zero entropy, brought by the $\bar{\rm p}$, and rank 2
because $\bar\alpha$ takes two values in (4.1).

An important simplification occurs due to the symmetry of reaction (1.1)
under the reflection $\Pi$ about the scattering plane,
which reverses the spin components parallel to this plane
(it is the ``B - symmetry" mentionned in~\cite{Doncel-M}).
Using the transversity basis, where $\zu \equiv \nv$, the scattering matrix $M$
is even under $ \sigma_x \to - \sigma_x $, $ \sigma_y \to - \sigma_y $,
and the amplitude (3.1) vanishes
when an odd number of transversities are negative.
Accordingly, the Grand Wave Function $\Psi$ has no components
like $ |\bar{u} \bar{u} u d \> $
and the original 4-particle density matrix $\R$
is reduced to $8 \times 8$.

For the same reason, the density matrices restricted to fewer particles like
$\R({\rm p}, \bar\Lambda, \Lambda)$, $\rho(\bar\Lambda, \Lambda)$ or $\rho(\Lambda)$
do not mix states with even and odd numbers of $d$'s.
Thus $\R({\rm p}, \bar\Lambda, \Lambda)$ is block-diagonal,
being the direct sum of two $4 \times 4$ matrices,
one corresponding to $\bar\alpha = \bar u $,
the other to $\bar\alpha = \bar d $ in Eq.~(4.1).
Each of these sub-matrices is of rank one.
Similarly, $\rho(\bar\Lambda, \Lambda)$ is block-diagonal
in two $2 \times 2$ sub-matrices of rank 2,
one corresponding to
$( \bar\alpha, \bar\beta )  = (\bar u, \bar u) $ or $ (\bar d, \bar d) $,
the other to
$( \bar\alpha, \bar\beta )  = (\bar u, \bar d) $ or $ (\bar d, \bar u) $
in Eq.~(3.12).
Finally, $\rho(\Lambda)$ is diagonal, which means that the $\Lambda$ polarization
is normal to the scattering plane.

\bigskip
\noindent
{\bf 5. The positivity conditions} 
\medskip

The positivity of the Grand Density Matrix comes from the very general,
but non-trivial requirement that the probability of any process is positive.
It is {\it not sufficient} to require that the cross section (2.1) is positive
for any set of polarizations $\left\{ \Sv_A,\Sv_B,\Sv_C,\Sv_D \right\}$.
Let us suppose, for instance, that (2.1) possesses the factor
$ (1 + \Sv_C \cdot \Sv_D) $.
This factor is positive or null for any $\Sv_C$ and $\Sv_D$.
However, it corresponds to a final density matrix of the form
$\rho_{C,D}  =  [ 1 + \sigma^i_C \otimes \sigma^i_D ]/4$
which is non-positive. For example, for the  {\it singlet} spin state,
we have $\sigma^i_C \otimes \sigma^i_D  = -3$.
The probability that reaction (1.1) produces a $(\bar\Lambda, \Lambda)$
system in the singlet state would be negative!
Note that the singlet state is {\it entangled}.
This shows that positivity has to be tested with non-entangled
{\it and} entangled states.

Similarly, a factor $ (1 - \Sv_A \cdot \Sv_C) $,
which leads to the complete spin reversal $\Sv_C = - \Sv_A$ according to (2.2),
gives a non-positive $R$ and is therefore forbidden.
As an example, let us consider the splitting $\pi \to q + \bar q$
followed by a quark--hadron scattering $q + h \to  q' + h'$
where the $\bar q$ is spectator.
The intermediate spin correlation  is
in $ (1 - \Sv_q \cdot \Sv_{\bar q}) $.
If there were a complete spin reversal $\Sv_q = - \Sv_{q'}$
in the quark-hadron scattering,
it would lead to a final correlation in $ (1 + \Sv_{q'} \cdot \Sv_{\bar q}) $,
which is forbidden as explained before.

The general positivity conditions are as follows:
a $N\times N$ hermitian matrix $ \rho $ is positive (respectively semi-positive)
if all its eigenvalues $r_i$ are positive (resp. positive or null).
Let us consider the symmetric functions of the eigenvalues
$$
\S_1 = \sum_{i} r_i \,,\quad
\S_2 = \sum_{i<j} r_i r_j \,,\quad
\S_3 = \sum_{i<j<k} r_i r_j r_k \,, \quad
\cdots \quad
\S_N = r_1 r_2 \cdots r_n~.
\eqno(5.2)
$$
$\S_n$ is the sum of the {\it on-diagonal sub-determinants} of order $n$
(when a sub-matrix has its diagonal on the diagonal of $ \rho $, we call it
"on-diagonal").
A necessary and sufficient condition of positivity,
or semi-positivity with $N_0$ vanishing eigenvalues, is
$$
\S_n > 0 \quad {\rm for} \quad  n \le N-N_0 \,,
\qquad
\S_n = 0 \quad {\rm for} \quad  N-N_0 < n \le N \,.
\eqno(5.3)
$$
If $ \rho $ is (semi-)positive, {\it each} of its on-diagonal sub-determinant
is (null or) positive. This may provide inequalities
simpler than, but redundant with (5.3),
in the same manner as $ |x^{2}| < 1 $ is redundant with $ |x^{2}| + |y^{2}| < 1 $.

The matrix $ \rho $ depends on $N^{2}$ real parameters.
They can be $ {\rm Re} (\rho_{ii'})$ for $i \le i'$
and $ {\rm Im} (\rho_{ii'})$ for $i < i'$,
or the correlation parameters, which are linear combinations of them.
In the $N^{2}$-dimensional parameter space,
the domain of positivity of $ \rho $ is a convex half-cone.
Its intersection with the hyperplane $ \S_1 \equiv \tr ( \rho ) = 1 $
is a finite convex domain $\D$.
The boundary of $\D$ is a sheet of the hypersurface $\S_N \equiv \det( \rho ) = 0$.
It is a $(N^{2} - 2)$-dimensional manifold of degree $N$.
On this boundary, $ \rho $ is only semi-positive.
The other conditions, $\S_n \ge 0$ for $n = 2, \cdots N-1$
define domains which {\it include} $\D$.
These ``auxiliary'' conditions serve to eliminate the other sheets of the
hypersurface $\S_N = 0$.
The hypersurface where $\S_n$, or any on-diagonal sub-determinant, vanishes is
externally tangent to $\D$.

As we have seen, for an exclusive reaction the Grand Density Matrix $\R$
is of rank one. Therefore all the $\S_n$'s are vanishing for $ n \ge 2 $
and all the positivity constraints are saturated.
It can be shown that $\R$ is on a "corner" of $\D$.
On the contrary, when much information is lost
through non-detected or non-analyzed particles, $\R$ is "deep inside" $\D$.

\bigskip
\noindent
{\bf 6. Empirical approach}
\medskip

The search for inequalities is straightforward using the density matrix
method, but does not reveal at once the shape of the allowed domains.
Also, when one writes the conditions on the density matrix, one gets in general a
combination of several spin observables, and thus one has to make appropriate
combinations of inequalities to obtain constraints on two or three given
observables of interest.

\begin{figure}[!!!!htbc]
\centering{%
\includegraphics[width=.99\textwidth]{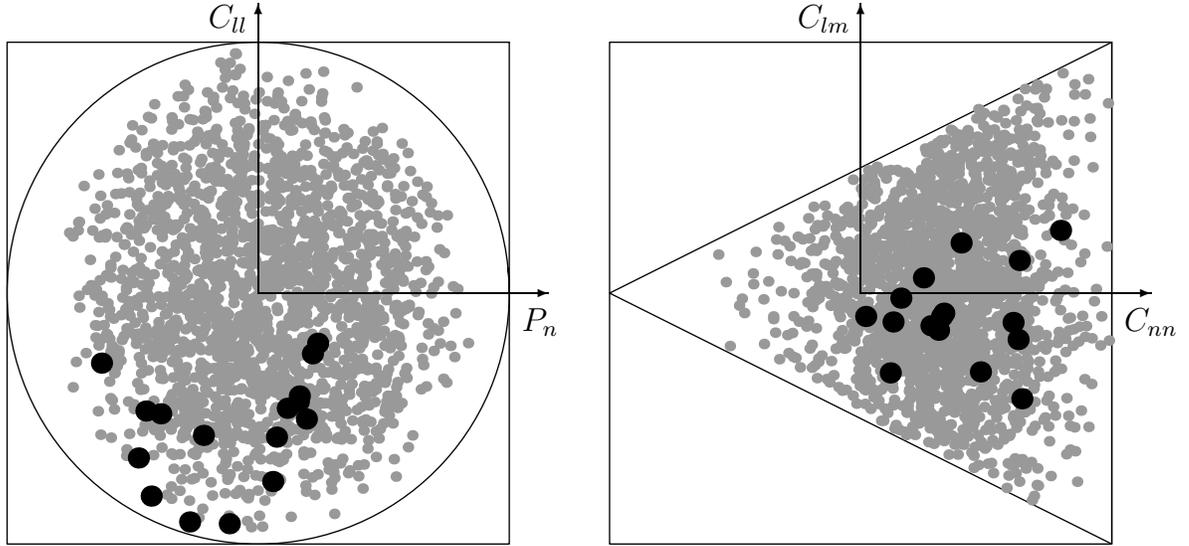}}
\caption{\label{Two} Pair of observables
restricted to the unit disk (left, here  polarisation and $C_{ll}$ are shown)
or to a triangle (right, here $C_{nn}$ and $C_{lm}$ are shown).
The small grey dots  correspond to hypothetical, randomly
generated, amplitudes, and the larger dots, to actual data.}
\end{figure}

\begin{figure}[!!!htbc]
\centering{%
\includegraphics[width=.99\textwidth]{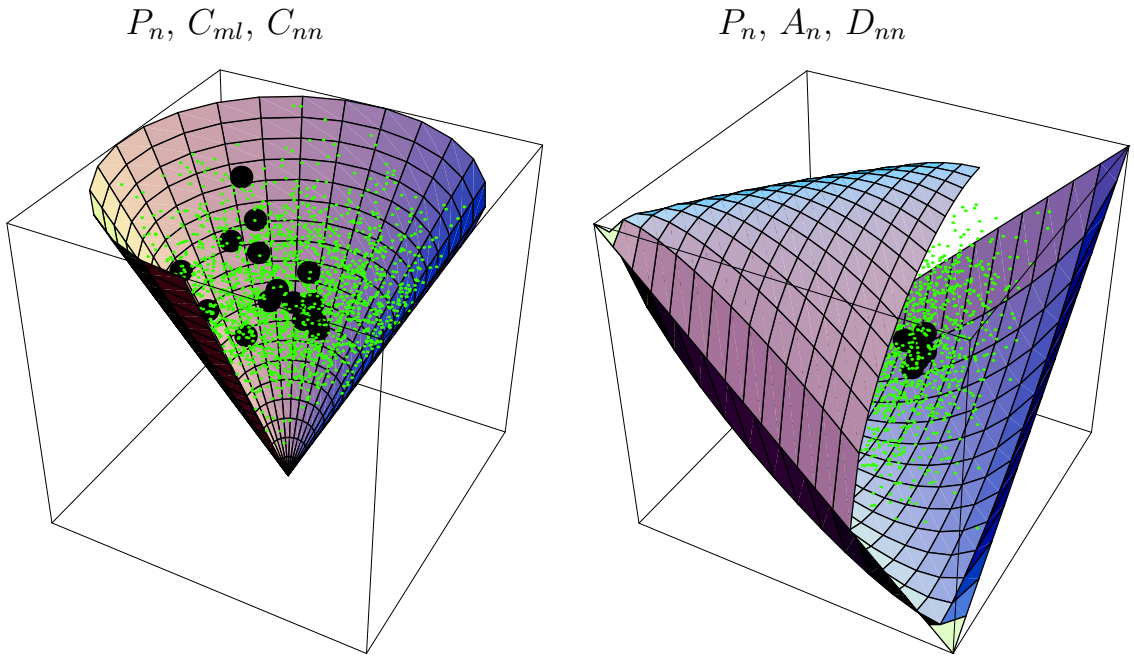}}
\caption{\label{Three} Triplet of observables restricted to the inner
volume a cone or of a cubic. The small dots correspond to randomly
generated amplitudes, the larger ones (partly hidden)  to actual data.}
\end{figure}

To circumvent this difficulty, the following method was
used in  Ref.~\cite{Elchikh}. The real and imaginary
parts of the amplitudes were chosen randomly, and the spin observables were
computed using their explicit expression in terms of amplitudes.  This detects
which pairs or triplets of
observables fulfill inequalities, and then these inequalities can be derived by
straightforward calculus. The case of pairs of observables is extensively
discussed in Ref.~\cite{Elchikh}, and preliminary results on triplets presented
at the LEAP2003 conference \cite{LEAP2003}. A sample of the results is
displayed in Figs.~\ref{Two} and \ref{Three}.

Notice that only a few types of inequalities are encountered.  For pairs of
observables, say $X$ and $Y$, each being typically restricted to $[-1,+1]$,
one gets the following possibilities
\begin{itemize}
\item nothing: $X$ and $Y$ might reach any point of the square $[-1,+1]^{2}$,
\item the disk $X^{2}+Y^{2}\le 1$,
\item a triangle $4Y^{2}\le(1+X)^{2}$.
\end{itemize}

For triplets of observables, say $X$, $Y$ and $Z$, the following situations are
obtained
\begin{itemize}
\item nothing, any point of cube $[-1,+1]^3$ is allowed,
\item a sphere $X^{2}+Y^{2}+Y^{2}\le 1$,
\item a cone $ (1+Z)^{2}\ge 4X^{2}+4Y^{2}$,
\item a cubic of the type $X^{2}+Y^{2}+Z^{2}\pm X Y Z \le 1$.
\end{itemize}
The latter case is the most interesting, since the domain is restricted in
space of three observables, but each projection cover the whole square,
i.e., there is no restriction for any pair observables within $(X,Y,Z)$. The
border has the shape of a twisted cushion.

\bigskip
\noindent
{\bf 7. Inclusive case: the spin-dependent parton densities}
\medskip

As an example of {\it inclusive} reaction,
let us consider now the elementary process (1.2),
which we rewrite for fixed momenta and spin vectors as
$$
N(\pv,\Sv_N) \to q(\kv,\Sv_q) + X \,,
\eqno(7.1)
$$
with $\kv = x \pv + \kv_T$.
The probability of (7.1) is the {\it spin- and $\vec k_T$-dependent}
quark density in the nucleon.
All what we will say below also applies to the quark fragmentation
$q \to$~baryon~+~X, only commuting $q$ and $N$,
or to any inclusive reaction of the type
$$
A\uparrow + B \to C\uparrow + X \,.
\eqno(7.2)
$$
already treated by Doncel and M\'endez~\cite{Doncel-M}.
Thus the problem was solved long ago.
One can also relate the inequalities in (7.2) to those of the
crossed reaction~\cite{Soffer-03}
$$
A_1\uparrow + A_2\uparrow  \to A_3 + X~,
\eqno(7.3)
$$
by the correspondence $ \Sv(A_2) \leftrightarrow - \Sv(C) $.

For a given spectator state $X$, the matrix element in spin Hilbert space
can be written as $\< \beta  | M_X | \alpha \>$.
Like in (3.6 - 3.7), we consider the fictitious crossed process
$$
\bar X \to \bar N(\pv,-\Sv_N) + q(\kv,\Sv_q) ~,
\eqno(7.4)
$$
where $ | \bar X \> \equiv \ CPT \ | X \> $.
Note that we have also moved the spectator system $X$ to the {\it initial} state.
The Grand Wave Function and Grand Density Matrix are then defined by
$$
\< \bar\alpha \beta | \Psi_X \>  =  \< \beta  | M_X| \alpha \> \,,
\eqno(7.5)
$$
$$
\R = \sum_X  | \Psi_X \>  \, \< \Psi_X | \,.
\eqno(7.6)
$$
Thus $\R$ corresponds to a statistical mixture.
Its rank $r$ is the dimension of the sub-space
spanned by the vectors $| \Psi_X \> $
in the ($\bar N q$) spin Hilbert space.
It cannot exceed the number of possible quantum states of the spectator system.
In general $r>1$, which means that some information is lost,
taken away by the spectator partons.

\def\oo{C_{00}}
\def\oz{C_{0z}}
\def\zo{C_{z0}}
\def\zz{C_{zz}}
\def\xx{C_{xx}}
\def\xy{C_{xy}}
\def\yx{C_{yx}}
\def\yy{C_{yy}}

In our case $R$ has dimension $4\times 4$ and depends
on {\it a piori} 16 correlation parameters $C_{\mu\nu}$ through the analog of (3.13).
However, like in the $2 \to 2$ reaction (1.1),
the plane defined by $\pv$ and $\kv_T$ is a symmetry plane.
It is therefore convenient to use the {\it transversity} basis
with $\zu$ normal to this plane, instead of the helicity basis
(unless one integrates over $\kv_T$).
In this basis $\R$ is even under
$ \sigma_x \to - \sigma_x $, $ \sigma_y \to - \sigma_y $
and the only non-vanishing coefficients are
$\oo \equiv 1$,
$\oz,$
$\zo,$
$\zz,$
$\xx,$
$\xy,$
$\yx$ and
$\yy$.
If we take $\xu$ along $\pv$ these parameters are respectively proportionnal to
$f_1$, $-h_1^\perp$, $f_{1T}^\perp$, $h_1-h_{1T}^\perp$,
$g_1$, $h_{1L}^\perp$, $g_{1T}$ and $h_1+h_{1T}^\perp$
of Ref.~\cite{Bacchetta-BHM},
all kinematical factors in $p_T/M$ or $p_T^{2}/(2M^{2})$ removed.
However the following inequalities are independent of the choice of the
$x$ and $y$ axes in the production plane.
Let us introduce
$$
{ 1 \pm \zz \over2} \equiv D_{nn}^ \pm \,,\quad
{  \oz  \pm \zo  \over2} \equiv A_n^ \pm \,,\quad
{  \xx \pm  \yy \over2} \equiv U^\pm \,,\quad
{  \xy  \pm  \yx \over2} \equiv V^\pm
\,.
\eqno(7.7)
$$
Putting the $ |\bar{N} q \> $ basic spin states in the order
$ \{   
|\bar{u} u \> ,\, |\bar{u} d \> ,\, |\bar{d} u \> ,\, |\bar{d} d \>
\} $, one has
$$
R = {1\over2} \ \pmatrix{
D_{nn}^+   +   A_n^+   &     0     &      0      &     U^+    - i V^+
\cr
0     &    D_{nn}^-   -   A_n^-     &   U^- + i  V^+     &      0
\cr
0     &   U^-  -i  V^+    &    D_{nn}^-   +   A_n^-      &      0
\cr
U^+   + i V^-    &     0     &    0    &    D_{nn}^+   -   A_n^+
\cr
}~.
\eqno(7.8)
$$
As expected from the symmetry about the $(x,y)$ plane,
$\R$ is block-diagonal in two rank-2 sub-matrices,
which obey the separate positivity conditions:
$$
( D_{nn}^\pm )^{2}  \ge (A_n^\pm )^{2}  + ( U^\pm )^{2}  + ( V^\mp )^{2}~,
\eqno(7.9)
$$
and
$$
D_{nn}^\pm   \ge 0 \,, \quad i.e., \quad  | \zz | \equiv |D_{nn}| \le 1 \,,
\eqno(7.10)
$$
which agrees with the results of Bacchetta {\it et al}~\cite{Bacchetta-BHM}
and of Ref.~\cite{Doncel-M}.

If we integrate over $\kv_T$, the only surviving parameters are
$\oo \equiv 1$, $\xx   \equiv \Delta q(x) / q(x)$
and  $\yy = \zz  \equiv \delta q(x) / q(x)$,
where $q(x)$, $ \Delta q(x) $  and $ \delta q(x) $ are the
quark {\it number}, quark {\it helicity}
and quark {\it transversity}~\cite{Barone,Artru-rhodan} distributions.
One obtains the Soffer inequality~\cite{Soffer}:
$$
2 \delta q(x) \le q(x) + \Delta q(x) \,.\eqno(7.11)
$$

In a simple model of quark distribution where $X$ just consists in a scalar di-quark,
all the inequalities (7.9 - 7.10) of the $\kv_T$ -dependent case are saturated.
Indeed, such an object has no spin to carry quantum information away;
a fully polarized nucleon delivers a fully polarized quark~\cite{Artru-rhodan}.
This is no more the case if we integrate over the degree of freedom $\kv_T$.
Nevertheless, the Soffer bound keeps saturated.

\bigskip
\noindent
{\bf 8. Conclusions and outlook}
\medskip

The formalism developed in the 60's remains extremely powerful to analyze
the consistency of spin observables. However, it needs some freshening
and new methods are needed to quickly derive the inequalities within a subset
of accessible observables.
We hope to have worked in this direction.

One of the basic tool, already used in~\cite{Doncel-M},
is a fictitious crossing which gathers all the particles on the same side.
It is expressed as a partial matrix transposition.
Usual crossing also links different physical reactions,
just reversing the polarization vectors.

Particle spin physics also touches the more general theory of quantum information,
in particular with the concept of entanglement.
The fact that the particles considered here have definite momenta is not essential.
The inequalities obtained in particle physics could also apply
to ``gates" between other kinds of quantum information
channels like optical fibers.

In a forthcoming article, we shall provide more details about the
positivity conditions and their physical interpretation.
In particular, we will show
explicitly that the method of the Grand Density Matrix gives the same
constraints on observables that the empirical approach based on randomly
generated amplitudes.

\bigskip
\noindent
{\bf Appendix A. Effect of crossing on the operators acting on an initial
particle}
\medskip

Together with (3.7), we have
$ \< \alpha' | \ \to \ | \bar\alpha' \> $,
$$
| \alpha \>  \< \alpha' | \ \to \
| \bar\alpha' \>  \< \bar\alpha | ~,
\eqno(\rm{A}.1)
$$
and for a linear combination of such elementary operators,
$$
\sum_{\alpha,\alpha'}  | \alpha \>  A_{\alpha\alpha'}  \< \alpha' |
\quad \to \quad \
\sum_{\alpha,\alpha'}  | \bar\alpha' \>  A_{\alpha\alpha'}  \< \bar\alpha | ~,
\eqno(\rm{A}.2)
$$
Here we have assumed that crossing acts {\it linearly} on operators.
Indeed (3.7) is the product of two anti-linear operations:
(i) applying CPT (ii) changing a ket into a bra.
Equation (A.2) amounts to the matrix transposition $A \to A^t$,
provided we choose the same ordering for the crossed basis vectors
$ \{ | \bar u \>  ,  | \bar d \> \} $
as in the initial basis vectors
$ \{ | u \>  , | d \> \} $
(the ordering in {\it magnetic} number $s_z$ is reversed:
it becomes $ \{ | \moins \>  , | \plus \> \} $).
For a single-spin matrix density, the transformation is
$$
\rho = {1\over 2} (1 + \Pv \cdot \sigma)
\quad \to \quad
\rho^t = \rho^\dagger = {1\over 2} (1 + \bar\Pv \cdot \underline{\sigma})~,
\eqno(\rm{A}.3)
$$
where $\bar\Pv = - \Pv$ due to spin reversal,
and $ \underline{\sigma_i} = - \sigma_i^t $ are the Pauli matrices
for the $\{ \bar 2 \}$ representation of SU(2)
(which is not often used, due to the equivalence
$\{ 2 \} \leftrightarrow \{ \bar 2 \}$).

\vspace{.3cm}
\noindent\emph{Acknowledgements.} We thank M. Elchikh for useful correspondence.
We also got experience in the properties of the positivity domain
in collaboration with E. Loyer~\cite{Loyer} during his academic training at IPNL.

%

\begin{thebibliography}{99}
%
\bibitem{Leader}
E.~Leader,
``Spin in Particle Physics,''
Cambridge Monogr.\ Part.\ Phys.\ Nucl.\ Phys.\ Cosmol.\  {\bf 15}
(2001) 1.
%
\bibitem{Klempt}
E.~Klempt, F.~Bradamante, Anna~Martin, J.-M.~Richard,
Phys.\ Rept.\  {\bf 368} (2002) 119.
%
\bibitem{Dolomieu} see, for instance, R. Omn\`es in
{\it Physics of Entangled States},
ed. by R. Arvieu and S. Weigert (Frontier Group, 2002), pp. 3-19;
J.I. Cirac, {\it ibid.} pp; 20-56.
%
\bibitem{Doncel-M}
M.G. Doncel and A. M\'endez, phys. Lett. 41B (1972) 83.
%
\bibitem{Elchikh}
M.~Elchikh and J.~M.~Richard,
Phys.\ Rev.\ C {\bf 61} (2000) 035205.
%
\bibitem{LEAP2003}
J.~M.~Richard and X.~Artru,
Proc. LEAP2003 Conference, Yokohama, Japan, 2003,
Nucl.\ Instrum.\ Meth.\ B {\bf 214} (2004) 171.
%
\bibitem{Soffer-03}
J. Soffer, Phys. Rev. Lett. 91 (2003) 092005.
%
\bibitem{Bacchetta-BHM}
A. Bacchetta, M. Boglione, A. Henneman, P.J. Mulders,
Phys. Rev. Lett. 85 (2000) 712.
%
\bibitem{Barone}
V. Barone, A. Drago, P.G. Ratcliffe, Phys. Rep. 359 (2002) 1-168.
%
\bibitem{Artru-rhodan} X.Artru, in {\it Spin in Physics},
ed. by M. Anselmino, F. Mila and J. Soffer, pp. 115-135 (Frontier Group, 2002)~;
ArXiv:hep-ph/0207309.
%
\bibitem{Soffer}
J. Soffer, Phys. Rev. Lett. 74 (1995) 1292.
%
\bibitem{Loyer} E. Loyer, {\it Corr\'elations de spins dans les collisions
\`a haute \'energie} (1997), unpublished.


%
\end{thebibliography}
\end{document}